\documentclass[11pt]{article}
\usepackage{amsmath}
\usepackage{amssymb}
\usepackage{amsthm}
\usepackage{amsfonts}
\usepackage{graphicx}
\usepackage{textcomp}
\usepackage{color}
\usepackage{hyperref}
\usepackage{slashed}
\usepackage{tikz-cd}
\numberwithin{equation}{section}

\begin{document}

\begin{titlepage}

\title{Kaluza-Klein Reduction of the 6 Dimensional \\ Dirac Equation on $\mathbb{S}^3 \cong SU(2)$ and  \\ Non-abelian Topological Insulators}

\author{Tekin Dereli\footnote{tdereli@ku.edu.tr}, Keremcan Do\u{g}an\footnote{kedogan@ku.edu.tr}, Cem Yeti\c{s}mi\c{s}o\u{g}lu\footnote{cyetismisoglu@ku.edu.tr} \\ \small Department of Physics, Ko\c{c} University, 34450 Sar{\i}yer, \.{I}stanbul, Turkey}

\date{}

\maketitle

\begin{abstract} \noindent In this work, the Kaluza-Klein reduction of the Dirac equation on a 6 dimensional spacetime $\mathbb{M}^{1+5} := \mathbb{M}^{1+2} \times \mathbb{S}^3$ is studied. Because of the group structure on $\mathbb{S}^3$, $\mathbb{M}^{1+5}$ can be seen as a principal $SU(2)$ bundle over the model Lorentzian spacetime $\mathbb{M}^{1+2}$. The dimensional reduction induces non-minimal $SU(2)$ couplings to the theory on $\mathbb{M}^{1+2}$. These interaction terms will be investigated by comparing with a minimally $SU(2)$ coupled Dirac equation on $\mathbb{M}^{1+2}$. We hope that these results may help us to understand non-abelian interactions of topological insulators.
\end{abstract}

\vskip 2cm


\thispagestyle{empty}

\end{titlepage}

\maketitle

\section{Introduction}

\noindent In recent years, topological insulators have been studied excessively by both theoretical \cite{kane-mele1}, \cite{kane-mele2} and experimental physicists \cite{konig-etal}. They are characterized by the non-trivial topological properties of its valence band bundle. It is a subbundle of the trivial vector bundle of Hilbert spaces, that Bloch Hamiltonians act on, over the base space that is the Brillouin zone torus \cite{frucharta-carpentiera}. They can be described by $s$ spatial dimensional bulk band structure and by bulk/boundary correspondence, they have $s-1$ dimensional surface modes which are typically described by Dirac Hamiltonians. Topological invariants of the bulk differentiates the phases of systems whose Hamiltonians have the same symmetry \cite{bernevig}. For example, an $s = 2$ dimensional topological insulator without time-reversal symmetry can be described by its first Chern number $\nu \in \mathbb{Z}$, and its quantized Hall conductivity is given by $\sigma = - \nu \frac{e^2}{2 \pi}$, where $e$ is the electron charge. By using the symmetry properties of these Dirac Hamiltonians, free topological insulators are classified in any dimension \cite{kitaev}. Some of the methods about this classification use dimensional reduction techniques \cite{ryu-etal}. Even though there are some progress \cite{ryu-takayanagi}, \cite{ryu-moore-ludwig}, the complete classification of interacting topological insulators is missing. We hope that our work can be useful for this classification. \\

\noindent In this paper, we consider the Kaluza-Klein reduction of the Dirac equation on a 6 dimensional spacetime $\mathbb{M}^{1+5} := \mathbb{M}^{1+2} \times \mathbb{S}^3$, where $\mathbb{M}^{1+2}$ is a 3 dimensional Lorentzian manifold with vanishing first and second Stiefel-Whitney classes and $\mathbb{S}^3$ is the 3-sphere. $\mathbb{S}^3$ can be endowed with $SU(2)$ group structure, so that $\mathbb{M}^{1+5}$ can be seen as a trivial principal $SU(2)$ bundle over $\mathbb{M}^{1+2}$. The group structure on $\mathbb{S}^3$ presents itself as minimal and non-minimal $SU(2)$ gauge couplings to the reduced theory on $\mathbb{M}^{1+2}$. We believe that these couplings may help us to understand the topological insulators in artificial gauge potentials. A method for the creation of artificial non-abelian $SU(2)$ couplings is developed in shaken optical latices by means of periodic forcing \cite{hauke-etal}. \\ 

\noindent The main idea of the dimensional reduction is that one embeds the model spacetime into a larger manifold equipped with a fibre bundle structure over it. Then one investigates the physical quantities on the larger manifold and deduce their contributions on the model spacetime. After some related ideas of Nordstr\"{o}m, dimensional reduction was used by Kaluza and Klein in order to unify the only known forces of that time: gravity and electromagnetism. Starting from a pure gravity theory in $1 + 4$ dimensions, they obtained the Einstein-Maxwell theory in $1 + 3$ dimensions. Although there were problems with their approach, generalization of their ideas about dimensional reduction for non-abelian Lie groups \cite{kerner} has been widely used in physics \cite{witten}. \\

\noindent  Before presenting our calculations, we outline the necessary mathematical structures in order to understand the Dirac equation and Kaluza-Klain reduction. In the second section, we discuss real Clifford algebras and spin groups over arbitrary spacetimes and give their properties \cite{ablamowicz-etal}, \cite{gilbert-murray} and \cite{lounesto}. In the third section, we start by explaining smooth bundle structures \cite{kobayashi-nomizu} and define the spin structure in terms of them \cite{bourguignon-etal}. We, then, discuss the construction of global spinor fields over arbitrary spacetimes. After that, in the section four, we shortly summarize the ideas behind Kaluza-Klein reduction. Then we reduce the Dirac equation on $\mathbb{M}^{1+5} = \mathbb{M}^{1+2} \times \mathbb{S}^3$ and discuss our results by comparing them with minimal $SU(2)$ couplings on $\mathbb{M}^{1+2}$. Similar calculations in the language of differential forms can be found in the literature. Dimensional reduction of different Yang-Mills couplings on $\mathbb{S}^3$ can be found in \cite{dereli-ucoluk}. Dirac spinors interacting with gravity and electromagnetism is considered as a $1+4$ dimensional geometry in \cite{dereli-tucker}. Reduction of spinors coupled to dilaton gravity can be found in \cite{adak-dereli}. \\

\noindent Here, we summarize the conventions and notations of our paper. We use Einstein's summation convention and natural units $\hbar = c = 1$. To separate quantities with frame indices on different spaces, we label them with uppercase Latin letters $A, B, \ldots = 0, 1, 2, 5, 6, 7$, lowercase Latin letters $a, b, \ldots = 0, 1, 2$, Greek letters $\alpha, \beta, \ldots = 5, 6, 7$ for $\mathbb{M}^{1+5}$, $\mathbb{M}^{1+2}$ and $\mathbb{S}^3$ respectively. These indices will be raised and lowered by the metric of its spacetime which is a non-degenerate, symmetric, covariant tensor of rank 2. Notations for terms and operations on each spacetime are collected in Table (\ref{Table1}) below:
\begin{table}[h]
	\centering
	\caption{Notations}
	\label{Table1}
	\begin{tabular}{llll}
		\hline
		\rule{0pt}{10.8pt}Term / Space           & $\mathbb{M}^{1+5}$ & $\mathbb{M}^{1+2}$ 	& $\mathbb{S}^3$ 						 	  		\\	\hline 
		\rule{0pt}{10.8pt}Coordinates			& $x^M$			 & $x^m$			& $x^{\mu}$	\\ 
		Orthonormal basis 1-forms     & 	$E_A$	 & $e_a$	& $e_{\alpha}$	 	\\
		Connection 1-forms       & $\Omega_{A B}$ & $\omega_{a b}$		& $\omega_{\alpha \beta}$ \\ 
		Gamma-matrices    & $\Gamma^A$	   & $\gamma^a$		& $\gamma^{\alpha}$	\\
		Lorentz generators   & $\Sigma^{A B}$ & $\sigma^{a b}$ & $\sigma^{\alpha \beta}$   \\
		Vector fields        & $\mathbb{X}_A$   & $X_a$	& $X_{\alpha}$	\\
		Hodge map       & $\#$ 	   & $*$ & $\star$	
	\end{tabular}
\end{table} \\

\noindent In this paper, exterior differential forms on a smooth manifold will be used extensively. Hence basic definitions and notations are summarized in the following paragraph. \\

\noindent Let $M$ be a smooth manifold. The set of all exterior differential forms on $M$, $\Lambda M := \oplus_{p = 0}^n \Lambda^p M$, forms a graded skew-commutative algebra over the space of smooth functions $C^{\infty} M$, with the wedge product (anti-symmetrized tensor product) $\wedge: \Lambda^p M \times \Lambda^q M \to \Lambda^{p + q} M$, where $\Lambda^p M$ is the set of exterior differential $p$-forms on $M$ and $n$ is the dimension of $M$. On a coordinate chart $U$ of $M$, a $p$-form can be written as 
\begin{equation}
\omega := \frac{1}{p!} \omega_{i_1 \ldots i_p} dx^{i_1} \wedge \ldots \wedge dx^{i_p},
\end{equation}
where $\{ \omega_{i_1 \ldots i_p} \} \subseteq C^{\infty} U$. On this chart, the exterior derivative $d: \Lambda^p M \to \Lambda^{p+1} M$ acts on $\omega$ as
\begin{equation}
d \omega := \frac{1}{p!} \frac{\partial}{\partial^{[i}} \omega_{i_1 \ldots i_p]} dx^i \wedge dx^{i_1} \wedge \ldots \wedge dx^{i_p},
\end{equation}
where [ ]-brackets denote the anti-symmetrization of indices. There are two more important operations acting on $\Lambda M$. One of them is the Hodge isomorphism $*$, provided that $M$ is endowed with a semi-Riemannian metric. It is a map $*: \Lambda^p M \to \Lambda^{n-p} M$, defined as
\begin{equation}
* \omega := \frac{1}{p! (n-p)!} \epsilon^{i_1 \ldots i_p}_{\ \ \ \ \ i_{p+1} \ldots i_n} dx^{i_{p+1}} \wedge \ldots \wedge dx^{i_n},
\end{equation}
where $\epsilon_{i_1 \ldots i_n}$ is the totally anti-symmetric Levi-Civita symbol with $\epsilon_{1 \ldots n} = 1$. The other one is the interior product $\iota_X$ with respect to a vector field $X$ on $M$. It is a map $\iota_X: \Lambda^p M  \to \Lambda^{p-1} M$ defined by
\begin{equation}
\iota_X \omega := \frac{1}{(p-1)!} X^{i_1} \omega_{i_1 \ldots i_p} dx^{i_2} \wedge \ldots \wedge dx^{i_p},
\end{equation}
where $X := X^i \frac{\partial}{\partial x^i}$ with $\{ X^i \} \subseteq C^{\infty} U$ on the coordinate chart $U$. \\

\noindent 

\newpage


\section{Clifford Algebras}
Let $\mathbb{R}^{t + s}$ be the $n = t + s$ dimensional real quadratic vector space with the quadratic form $Q_{t+s}:\mathbb{R}^{t + s}\to \mathbb{R}$
\begin{equation} Q_{t+s}(u) := - \left( u_1^2 + \ldots + u_t^2 \right) + \left( u_{t+1}^2 + \ldots + u_n^2 \right),
\end{equation}
where $u := (u_1, \ldots, u_n) \in \mathbb{R}^n$. Let $\{e_i\}$ be a basis of $\mathbb{R}^{t + s}$ such that $B_{ij}:=B_{t+s}(e_i,e_j)=diag(-1_t,+1_s)$, where $B_{t+s}:\mathbb{R}^{t + s}\times\mathbb{R}^{t + s}\to\mathbb{R}$ is the bilinear form defined by the polarization identity
\begin{equation} B_{t+s}(u, v) := \frac{1}{2} \left[ Q(u) + Q(v) - Q(u - v) \right].
\end{equation}
A Clifford algebra $(A, \nu)$ over $\mathbb{R}^{t + s}$ is a unital associative algebra $A$ generated by the field $\mathbb{R}$ and the quadratic space $\mathbb{R}^{t + s}$ under the linear embedding
\begin{equation}
\nu: \mathbb{R}^{t + s} \to A, \qquad 1 \mapsto 1_A, \qquad e_i \mapsto \Gamma_i, 
\end{equation}
subject to the anti-commutation relation
\begin{equation}
\{ \Gamma_i, \Gamma_j \} := \Gamma_i \Gamma_j + \Gamma_j \Gamma_i = 2 B_{i j} 1_A,
\end{equation}
where $1_A$ denotes the unity in $A$\footnote{A Clifford algebra can be defined with respect to a more general quadratic form over a vector space.}. \\

\noindent In physics, the Clifford algebras that one encounters is the unique universal Clifford algebra. The universal Clifford algebra over $\mathbb{R}^{t + s}$ is denoted by $\left( Clif(t, s), \mu \right)$, or simply $Clif(t, s)$. Universality means that for every Clifford algebra $(A, \nu)$ over $\mathbb{R}^{t + s}$, there is a unital algebra homomorphism $\beta: (Clif(t, s), \mu) \to (A, \nu)$ such that the diagram
\begin{center}
\begin{tikzcd}
\mathbb{R}^{t + s} \arrow[r, "\mu"] \arrow[dr, "\nu"]
& Clif(t, s) \arrow[d, dashrightarrow, "\beta"] \\
& A
\end{tikzcd}
\end{center}
commutes. For every quadratic space, there is a universal Clifford algebra and it is unique up to unital algebra isomorphism. Moreover, a Clifford algebra is universal if and only if it is $2^n$ dimensional as a real vector space. The generators $\{ \Gamma_i \}$ of $Clif(t, s)$ are called gamma-matrices of $\mathbb{R}^{t + s}$. $Clif(t, s)$ is canonically isomorphic to the exterior algebra $\Lambda \mathbb{R}^{t + s}$ via
\begin{equation}
1_{Clif(t,s)} \mapsto 1, \qquad \Gamma_i \mapsto e_i, \quad \dots, \qquad  \Gamma_{i_1 \dots i_n} \mapsto e_{i_1} \wedge \dots \wedge e_{i_n},
\end{equation}
where the product of two generators are given by
\begin{equation}
\Gamma_i\Gamma_j= \frac{1}{2}[\Gamma_i,\Gamma_j] + \frac{1}{2}\{\Gamma_i,\Gamma_j\}=:\Gamma_{ij}+B_{ij}1_{Clif(t, s)}.
\label{prod}
\end{equation}
\noindent Clifford algebra $Clif(t, s)$ is intimately related to the orthogonal group $O(t, s)$. $O(t, s)$ acts on $Clif(t, s)$ via automorphisms. The orthogonal transformation $e_i \mapsto -e_i$ induces an automorphism whose action turns $Clif(t, s)$ into a $\mathbb{Z}_2$-graded algebra:
\begin{equation}
Clif(t, s)=Clif_0(t, s)\oplus Clif_1(t, s),
\end{equation} 
where $Clif_0(t, s)$ and $Clif_1(t, s)$ are spanned by products of even and odd number of basis elements respectively. \\

\noindent The even subalgebra $Clif_0(t, s)$ is closely related to special orthogonal group $SO(t, s)$ and its double cover $Spin(t, s)$, namely the spin group. Equipped with the Lie commutator, $Clif(t, s)$ is a Lie algebra. Because, the special orthogonal Lie algebra $\mathfrak{so}(t, s)$ is isomorphic to the space of 2-forms $\Lambda^2 \mathbb{R}^{t + s}$, it is Lie a subalgebra of $Clif(t, s)$ under the injective homomorphism:
\begin{equation}
e_i \wedge e_j \mapsto  \frac{1}{4}[\Gamma_i, \Gamma_j]=\frac{1}{2}\Gamma_{ij}, \quad i\neq j,
\end{equation}
where, $e_i \wedge e_j$ is the image of rotation generator in $\mathfrak{so}(t, s)$ which infinitesimally rotates the plane spanned by the vectors $e_i$ and $e_j$. Exponentiating $\mathfrak{so}(t, s)$ in $Clif(t, s)$ yields the identity component of spin group $Spin(t, s)$. There exists a short exact sequence of groups given by 
\begin{equation}
1 \longrightarrow \mathbb{Z}_2 \longrightarrow Spin(t, s) \longrightarrow SO(t, s) \longrightarrow 1.
\end{equation}
The homomorphism $\tilde{\varphi}: Spin(t, s) \to SO(t,s)$ is a double covering with kernel $\mathbb{Z}_2$. \\

\noindent The Clifford algebras are semi-simple algebras and by Wedderburn's theorem, are represented by finite dimensional matrix algebras over the normed division algebras $\mathbb{R}$, $\mathbb{C}$ or $\mathbb{H}$. The representations are classified by the dimension $n = t + s$ and the signature $s - t$ up to Morita equivalence. A spinor is defined as an element of the representation space of a faithful, irreducible projective representation of $Spin(t, s)$. 


\section{Spin and Spinor Bundles}

\noindent In order to define spin bundles, we need to discuss principal fibre bundles and their associated bundles. These geometric structures have ubiquitous applications in physics. \\

\noindent A smooth bundle is a triple $(P,\pi,M)$ where $P$ (called the total space) and $M$ (called the base space) are smooth manifolds and $\pi:P \to M$ is a smooth surjective map. Moreover, it should be locally trivial, i.e. for every open chart $U$ of $M$, there exists a diffeomorphism $\phi_U: \pi^{-1}(U) \to U \times F$ for some typical fiber $F$ which is also a smooth manifold. The smooth bundle is called a vector bundle if $F$ is a vector space and the local trivialization maps are linear isomorphisms of vector spaces. The smooth bundle $(P,\pi,M)$ is called a principal $G$ bundle if there exists a smooth right action $\triangleleft: P \times G \to G$ of a Lie group $G$ (called the structure group) on $P$, which preserves the fibers $P_m := \pi^{-1}(\{ m \}), m \in M$ and acts freely and transitively on them. Moreover, $g \mapsto p \triangleleft g$ is a diffeomorphism for all $m \in M, p \in P_m$ between $G$ and $P_m$. The action is transitive on the fibers, so $M$ is homeomorphic to the space of orbits $P / G$ and we have a principal bundle isomorphism
\begin{center}
\begin{tikzcd}
P \arrow[d, "\pi"]  \\
M 
\end{tikzcd}
\ $\cong$ \ 
\begin{tikzcd}
P \arrow[d, "\tilde{\pi}"] \\
P/G,
\end{tikzcd}
\end{center}
where $\tilde{\pi}: P \to P/G$ is the quotient map. The principal bundle isomorphism consists of two diffeomorphisms $(H,h)$ such that the diagram
\begin{center}
\begin{tikzcd}
P \arrow[r, "H"] \arrow[d, "\pi"]
& P \arrow[d, "\tilde{\pi}"] \\
M \arrow[r, "h"]
& P/G
\end{tikzcd}
\end{center}
commutes. The principal $G$ bundle $\pi:P\to M$ is locally a product space of the base manifold $M$ and the manifold $G$. Since the right $G$ action is free on the fibers, they have a $G$-torsor structure which means that they carry the group structure without a specified identity element. \\

\noindent Given a principal $G$ bundle $\pi:P\to M$, and a vector space $F$ equipped with a smooth left $G$ action $\triangleright: G\times F\to F$, we define an associated vector bundle $\pi_F: P_F \to M$ as follows: Let $P_F:= (P\times F)/\sim$ where $(p,f) \sim (p',f')$ if and only if there exists $g \in G$ such that $p' = p \triangleleft g $ and $f'=g ^{-1} \triangleright f$. Also, the projection map is defined as
\begin{align}
\pi_F: P_F &\to M \nonumber\\
[p,f] &\mapsto \pi(p).
\end{align}  
where $[p, f]$ is the equivalence class containing $(p, f)$. The associated bundle $\pi_F: P_F \to M$ is a vector bundle with base $M$ and typical fibre $F$ carrying a representation of the Lie group $G$. \\ 

\noindent In the context of spin bundles, one starts with the base manifold $M$ of dimension $n$ which models spacetime. One constructs the linear frame bundle $\pi:LM \to M$ of $M$ where 
\begin{equation}
LM:=\bigsqcup_{m\in M} L_mM
\end{equation}
and elements of $L_mM$ are the frames spanning the tangent space $T_mM$, that is $L_mM:=\big\{ \{ e_1,\dots,e_n \} \ | \ T_mM=span\{e_1,\dots,e_n\} \big\}$. The general linear group $GL(n,\mathbb{R})$ has a smooth right action on the frames given by an element $g = \left( g^i_{\ j} \right) \in GL(n, \mathbb{R})$:
\begin{equation}
e_j \triangleleft g := g^i_{\ j} e_i, \quad \text{for each $j \in \{1, \ldots, n \}$}.
\end{equation}
Hence the linear frame bundle is a principal bundle whose structure group is $GL(n, \mathbb{R})$. This construction is always possible over $M$ without any extra structure. If $M$ can be endowed with a semi-Riemannian metric $g$ of signature $s-t$, then we can construct the $g$-orthonormal frames. Considering the bundle of $g$-orthonormal frames $\pi: OM \to M$, reduces the structure group from $GL(n,\mathbb{R})$ to $O(t,s)$. Furthermore, if the manifold's first Stiefel-Whitney class $w_1(M) \in H^1 \left( M; \mathbb{Z}_2 \right)$ vanishes, then it is orientable. If $M$ is orientable, we can consider the bundle of oriented orthonormal frames $\pi:OOM \to M$ where the structure group further reduces to $SO(t,s)$. \\

\noindent In order to be able to discuss the spin frame bundle or shortly spin bundle on $M$, there is a topological obstruction similar to the orientability of $M$; the second Stiefel-Whitney class $w_2(M) \in H^2 \left( M; \mathbb{Z}_2 \right)$ of $M$ should vanish. If $M$ is spinnable, then the spin bundle\footnote{Inequivalent spin structures on $M$ are in one-to-one correspondence with the cohomology classes in $H^1 \left( M; \mathbb{Z}_2 \right)$.} \footnote{In particular, $\mathbb{S}^3$ has vanishing first and second Stiefel-Whitney classes. $H^1 \left( \mathbb{S}^3; \mathbb{Z}_2 \right)$ is trivial, so that it admits a unique spin structure. Additionally we assumed that $\mathbb{M}^{1+2}$ also admits a spin structure. Hence, a spin structure and spinor fields can be defined on the product manifold $\mathbb{M}^{1+5}$ \cite{milnor-stasheff}.} of $M$ is defined as a principal $Spin(t,s)$ bundle $\pi': SM \to M$ together with a principal bundle morphism $\left( \varphi, id_M \right)$ to $\pi:OOM\to M$ given as the commutative diagram
\begin{center}
\begin{tikzcd}
SM\arrow[r, "\varphi"] & OOM  \\
SM \arrow[u, "\triangleleft' "] \arrow[r, "\varphi"] \arrow[d, "\pi'"] & OOM \arrow[u, "\triangleleft"] \arrow[d, "\pi"] \\
M \arrow[r, "id_M"] & M,
\end{tikzcd}
\end{center}
where $\triangleleft': SM\times Spin(t,s) \to SM$ and $\triangleleft: OOM \times SO(t,s) \to OOM$ are the smooth right actions on total spaces and the equivariant map $\varphi: SM \to OOM$ restricts fibrewise to the covering homomorphism $\tilde{\varphi}: Spin(t,s) \to SO(t,s)$, i.e. $\phi(p \triangleleft' g) = \phi(p) \triangleleft \tilde{\phi}(g)$ for $p \in SM, g \in Spin(t, s)$. \\

\noindent Then the spinor bundle of $M$ is an associated vector bundle $\pi'_{\mathcal{H}}:SM_\mathcal{H} \to M$ to the spin bundle with typical fibre $\mathcal{H}$ equipped with a smooth left action $\triangleright':Spin(t,s) \times \mathcal{H} \to \mathcal{H}$. $\mathcal{H}$ is a Hilbert space that carries a faithful, unitary, projective representation of the spin group $Spin(t, s)$. Sheaf of smooth sections $\Gamma \left( SM_F \right) := \big\{ \psi: M\to SM_F \ | \ \pi'_F \circ \psi = id_M, \mbox{smooth} \big\}$ of this bundle consists of the spinor fields on $M$. In physics, one usually uses a local patch of coordinates defined only on a coordinate chart $U \subset M$. This yields local sections $\Gamma \left( SM_F, U \right)$ of $SM_F$ over the set $U$. Then, one can glue the local spinor fields on the intersections under the sheaf or locality condition to get one global spinor field over $M$. A spinor field $\psi$ on a semi-Riemannian manifold $M$ with the metric signature $s - t$ satisfies the Dirac equation, which can be expressed in a coordinate invariant way as:
\begin{equation}
\slashed{D} \psi = m * \psi,
\end{equation}
where $\slashed{D}$ is the Dirac operator on $M$ and $m \in \mathbb{R}^{\geq 0}$ is the mass of the spinor field. It is a first order differential operator of the form
\begin{equation}
\slashed{D} := * \tilde{\gamma} \wedge D,
\end{equation}
where $\tilde{\gamma} := \tilde{\gamma}^{a} \tilde{e}_{a}$ is the gamma-matrix valued 1-form given by the orthonormal basis 1-forms $\{ \tilde{e}_{a} \}$ and gamma-matrices $\{ \tilde{\gamma}^{a} \}$ of $M$. Gamma-matrices of a manifold are given in terms of the smooth sections of the Clifford bundle $Clif(M)$,
\begin{equation}
Clif(M) := \bigsqcup_{m \in M} Clif \left( T_m M, Q_{t+s} \right),
\end{equation}
where $Clif \left( T_m M, Q_{t+s} \right)$ is the Clifford algebra of $T_m M$ generated by the quadratic form $Q_{t+s}$\footnote{Locally $Clif(M)$ is just $Clif(t, s)$.}. $D$ is a $Spin(t, s)$ exterior covariant derivative operator given by
\begin{equation}
D = d + \frac{1}{2} \tilde{\omega}_{a b} \tilde{\sigma}^{a b},
\end{equation}
where $\{\tilde{\omega}_{{a} {b}}\}$ are the spin connection  1-forms\footnote{A connection on a principal $G$ bundle $\pi: P \to M$ is the $G$-invariant assignment of the horizontal subspace $H_p P$ for each tangent space $T_p P$ so that $T_p P = H_p P \oplus V_p P$, where $V_p P$ is the vertical subspace consisting of tangent vectors to the fibre. For a matrix Lie group $G$, it can be given by a set of local $\mathfrak{g}$ valued 1-forms $A_U \in \Gamma (M, U, \mathfrak{g})$ with the requirement 
\begin{equation}
A_V = g_{U V}^{-1} A_U g_{U V} + g_{U V}^{-1} d g_{U V},
\end{equation}
where $g_{U V}: U \cap V \to G$ is the transition function between the local trivialization maps on open sets $U$ and $V$. Each vector field can be uniquely decomposed into horizontal and vertical parts by the projection maps $\mathcal{H}$ and $\mathcal{V}$ respectively. The exterior $G$ covariant derivative associated with a connection $A$ is the linear operator $D_A:= \mathcal{H}^* \circ d: \Lambda^p \left( P, \mathfrak{g} \right) \to \Lambda^{p+1} \left( P, \mathfrak{g} \right)$, where $\Lambda^p \left( P, \mathfrak{g} \right)$ is the set of $\mathfrak{g}$ valued $p$-forms on $P$.} and $\{\tilde{\sigma}^{{a} {b}}\}$ are the Lorentz generators on $M$defined as $\frac{1}{4} \left[ \tilde{\gamma}^a, \tilde{\gamma}^b \right]$. \\


\section{Kaluza-Klein Reduction}

\noindent Einstein's general relativity explains gravity from a purely geometric perspective, a $1 + 3$ dimensional Lorentzian manifold $M$, modeling the spacetime, and an energy-momentum tensor which set to be equal to the Einstein tensor of $M$. The field equations and conservation laws follow from the Bianchi identity. To write Maxwell's electromagnetic theory on this spacetime, one has to use extra tools other than Lorentzian manifold's geometry. One needs to introduce a $\mathfrak{u}(1)$ valued closed differential 2-form on $M$ which represents the electromagnetic field. Two of the Maxwell's equations follow from $d^2 = 0$ (or again from the Bianchi identity) and the other two should be postulated in terms of a current 1-form. \\

\noindent Kaluza came up with the idea of a $1 + 4$ dimensional theory that combines gravity and electromagnetism. Later, Klein added the quantum interpretation and obtained that the extra dimension should be a circle of radius comparable to Planck length (order of $10^{-35}$ meter). By starting from a purely geometric theory for $\mathbb{R}^{1 + 3} \times \mathbb{S}^1$, they obtained the gravitational and electromagnetic interactions on the $1 + 3$ dimensional Minkowski spacetime $\mathbb{R}^{1 + 3}$ that we observe to live in. The theory was problematic and one of the solution ideas was to introduce a scalar potential known as Jordan-Thiry field which scales the extra dimensions. Nevertheless, the idea of Kaluza-Klein reduction outreached its first aim \cite{kerner} and even today its generalizations are a part of the current research \cite{witten}. \\

\noindent Electromagnetic field can be regarded as the curvature of a connection\footnote{The curvature of a connection $A$ of a principal $G$ bundle $\pi: P \to M$ is a $\mathfrak{g}$ valued 2-form defined as $F := D_A A$.} on a principal $U(1)$ bundle over the spacetime and $U(1)$ is diffeomorphic to the circle $\mathbb{S}^1$ as a manifold. In this sense, Kaluza-Klein theory considers a $1 + 4$ dimensional Lorentzian manifold $P$ equipped with a principal $U(1)$ bundle structure on $M$ where $M := P / \mathbb{S}^1$ is a $1 + 3$ dimensional manifold which model the extended spacetime. \\

\noindent This picture can be generalized for any Lie group $G$. In order to perform a general dimensional reduction, one needs to consider a principal $G$ bundle $\pi: P \to M$, where $M$ is a semi-Riemannian manifold with a metric $g$. Given a connection (or a gauge potential) $A$ on $P$ and an $\mathfrak{ad}$-invariant metric $h$\footnote{It is a non-degenerate bilinear form on the Lie algebra $\left( \mathfrak{g}, [\cdot, \cdot] \right)$ that satisfies 
\begin{equation} h \left( [x, y], z \right) + h \left( y, [x, z] \right) = 0, \qquad \mbox{for all $x, y, z \in \mathfrak{g}$.} \nonumber
\end{equation}}
on $\mathfrak{g}$, one can define a bundle metric $g_P$ as

\begin{equation}
g_P := \pi^* g + h_A,
\end{equation}
where $h_A (X, Y) := h \left( A(X), A(Y) \right)$ for all vector fields $X, Y \in \Gamma \left( T P \right)$ and $\pi^* g$ is the pullback of $g$ by $\pi$, i.e. $\pi^* g (X, Y) := g \left( \pi_* X, \pi_* Y \right)$, where $\pi_*$ is the pushforward of $\pi$ defined by the commutative diagram:
\begin{center}
\begin{tikzcd}
T P \arrow[r, "\pi_*"] \arrow[d]
& T M \arrow[d] \\
P \arrow[r, "\pi"]
& M.
\end{tikzcd}
\end{center}
Next, one chooses the action density as the scalar curvature of the bundle metric $g_P$. Variation of this action density with respect to the metric $g$ gives the Einstein field equations on $M$. Moreover, variation with respect to the connection $A$ leads to the Yang-Mills equations for the curvature of the connection (or the field strength of the gauge field) \cite{bleecker}. \\

\noindent This dimensional reduction technique can be applied directly to the field equations, in particular to the Dirac equation. We will be working with the Dirac equation on a 6 dimensional manifold $\mathbb{M}^{1+5} = \mathbb{M}^{1+2} \times \mathbb{S}^3$, where $\mathbb{M}^{1+2}$ is a Lorentzian manifold\footnote{A smooth manifold $M$ can be endowed with a Lorentzian metric if and only if $M$ is compact or non-compact with vanishing Euler characteristic 
\[ \chi(M) = \sum_{i} (-1)^i rank(H^i(M; \mathbb{Z})). \]} whose first and second Stiefel-Whitney classes vanish and $\mathbb{S}^3$ is the 3-sphere whose embedding in $\mathbb{R}^4$ is given by 
\begin{equation}
(y^1)^2 + (y^2)^2 + (y^3)^2 + (y^4)^2 = 1.
\end{equation}

\noindent $\mathbb{M}^{1+5}$ can be seen as a trivial principal $SU(2)$ bundle over $\mathbb{M}^{1+2}$ because $\mathbb{S}^3$ can be equipped with an $SU(2)$ Lie group structure via the maps $\mathbb{S}^3 \leftrightarrow S \mathbb{H} \to SU(2)$:
\begin{equation}
\left( y^1, y^2, y^3, y^4 \right) \mapsto y^1 + y^2 i + y^3 j + y^4 k \mapsto \left( \begin{matrix} y^1 + i y^2 & - y^3 + i y^4 \\ y^3 + i y^4 & y^1 - i y^2 \end{matrix} \right).
\end{equation} 
After the identification of $\mathbb{S}^3$ with the unit quaternions $S \mathbb{H}$ via the first map, the group multiplication on $\mathbb{S}^3$ is the quaternion multiplication and the inverse is the quaternion conjugation. These operations are smooth so that it becomes a Lie group and the second map establishes a Lie group isomorphism between $\mathbb{S}^3$ and $SU(2)$. $SU(2)$ is the double cover of the 3 dimensional rotation group $SO(3)$ and the Lie algebra $\mathfrak{su}(2)$ is a 3 dimensional vector space equipped with a Lie bracket given by 
\begin{equation}
\left[ X_{\alpha}, X_{\beta} \right] = \epsilon_{\alpha \beta}^{\ \ \ \gamma} X_{\gamma}.
\end{equation}
where $\{\epsilon_{\alpha \beta}^{\ \ \ \gamma} \}$ are the structure constants of $\mathfrak{su}(2)$ Lie algebra\footnote{It is a metrisable Lie algebra, i.e. one can find an $\mathfrak{ad}$-invariant metric on it. There is a one-to-one correspondence between $\mathfrak{ad}$-invariant metrics of Lie algebras and bi-invariant metrics of the corresponding Lie group. A metric $\left< \cdot, \cdot \right>$ on a Lie group $G$ is called left-invariant if 
\begin{equation}
\left< X, Y \right> = \left< L_{g *} X, L_{g *} Y \right>, \qquad \mbox{for all vector fields $X, Y \in \Gamma (T G)$},
\end{equation}
where $L_{g}$ is the left action of the group on to itself by an element $g \in G$. A metric which is both left- and right-invariant is called bi-invariant. $SU(2)$ is compact, so negative of the Cartan-Killing form $Tr \left[ ad_x ad_y \right]$ defines an $\mathfrak{ad}$-invariant metric on $\mathfrak{su}(2)$, where $ad_x(y) := [x, y],$ for $x, y \in \mathfrak{g}$. This induces a corresponding bi-invariant metric on $SU(2)$ which is unique up to scaling. $SU(2)$ with this metric becomes isometric to $\mathbb{S}^3$.}. By using the metrics on $\mathbb{M}^{1+2}$ and $\mathbb{S}^3$ with given $SU(2)$ connection 1-forms $\{ A_{\alpha} := A_{\alpha}^{\ \ a} e_{a} \}$, one can construct the bundle metric $g_{\mathbb{M}^{1+5}}$ on $\mathbb{M}^{1+5}$:
\begin{equation} g_{\mathbb{M}^{1+5}} = \eta^{A B} E_A \otimes E_B,
\end{equation}
where $\eta^{A B} = diag(- + + + + +)$. The orthonormal basis 1-forms $\{ E_A \}$ of $\mathbb{M}^{1+5}$ are given by the Kaluza-Klein ansatz:
\begin{align} E_a &= e_a(x^m), \nonumber \\
							E_{\alpha} &= e_{\alpha}(x^{\beta}) + A_{\alpha}(x^m),
\label{ansatz}
\end{align}
where $\{ e_{a} \}$ and $\{ e_{\alpha} \}$ are the orthonormal basis 1-forms on $\mathbb{M}^{1 + 2}$ and $\mathbb{S}^3$ respectively. The Dirac equation on $\mathbb{M}^{1+5}$ is given by
\begin{equation} \# \Gamma \wedge D \Psi = M \# \Psi ,
\label{e10}
\end{equation}
where $\Psi$ is a spinor field on $\mathbb{M}^{1+5}$ and $M \in \mathbb{R}^{\geq 0}$ is its mass. Moreover, $\Gamma := \Gamma^A E_A$ is the gamma-matrix valued 1-form on $\mathbb{M}^{1+5}$ and $D := d + \frac{1}{2} \Omega_{A B} \Sigma^{A B}$ is the $Spin(1,5)$ exterior covariant derivative. \\

\noindent By using the ansatz (\ref{ansatz}), we can express the quantities on $\mathbb{M}^{1+5}$ in terms of the ones on $\mathbb{M}^{1+2}$ and $\mathbb{S}^3$. For instance, the interior product operation on $\mathbb{M}^{1+5}$ decompose as follows:
\begin{align} \mathbb{\iota}_{\mathbb{X}_A} &= \iota_{X_a} - A^{\alpha}_{\ a} \iota_{X_{\alpha}}, \nonumber \\ 
							\mathbb{\iota}_{\mathbb{X}_{\alpha}} &= \iota_{X_{\alpha}}.
\label{e2}
\end{align}
Assuming a torsion-free Levi-Civita connection on $\mathbb{M}^{1+5}$, the connection 1-forms $\{ \Omega_{A B} \}$ can be solved algebraically from the Cartan's first structure equation
\begin{equation}
dE_A+\Omega_A^{\ \ B} \wedge E_B = 0,
\end{equation}
using
\begin{equation} 
\Omega_{A B} = \frac{1}{2} \Big[ - \iota_{\mathbb{X}_A} \left( d E_B \right) + \iota_{\mathbb{X}_B} \left( d E_A \right) + \iota_{\mathbb{X}_A} \iota_{\mathbb{X}_B} \left( d E_C \right) E^C \Big]. \label{e3}
\end{equation}
Connection 1-forms are decomposed as follows:
\begin{align}	\Omega_{a b} &= \omega_{a b} - \frac{1}{2} F^{\gamma}_{\ a b} \left( e_{\gamma} + A_{\gamma} \right), \nonumber \\ 
\Omega_{a \beta} &= - \Omega_{\beta a} = - \frac{1}{2} F_{\beta a}^{\ \ b} e_b, \nonumber \\ 
\Omega_{\alpha \beta} &= \omega_{\alpha \beta} - \frac{1}{2} \epsilon_{\alpha \beta}^{\ \ \ \gamma} A_{\gamma} = \frac{1}{2} \epsilon_{\alpha \beta}^{\ \ \gamma} \left( e_{\gamma} - A_{\gamma} \right),
\label{e4}
\end{align}
where $\{ F_{\alpha} = d A_{\alpha} + \frac{1}{2} \epsilon_{\alpha}^{\ \ \beta \gamma} A_{\beta} \wedge A_{\gamma} = \frac{1}{2} F_{\alpha}^{\ b c} e_b \wedge e_c \}$ are the $SU(2)$ curvature 2-forms. Also, we made use of the Cartan-Maurer structure equation 
\begin{equation}
de_a=\epsilon_{a}^{\ b c}e_{b} \wedge e_{c},
\label{CM}
\end{equation}
in the last equality of (\ref{e4}) for reading the torsion-free connection 1-forms $\{ \omega_{\alpha\beta} \}$ on $\mathbb{S}^3$. \\

\noindent The Hodge duals of orthonormal basis 1-forms $\{ E^A \}$ decompose as:
\begin{align} 
\# E_{a} &= * e_{a} \wedge \star 1 + * 1 \wedge  A^{\gamma}_{\ a} \star e_{\gamma}, \nonumber \\ 
							\# E_{\alpha} &= - * 1 \wedge \star e_{\alpha}, \nonumber \\ 
							\# 1 &= * 1 \wedge \star 1.
\label{e5}
\end{align}
For constructing the gamma-matrices of $\mathbb{M}^{1+5}$, we use the Clifford algebras $Clif(1,2)$ and $Clif(0,3)$ of $\mathbb{R}^{1+2}$ and $\mathbb{R}^{0+3}$ respectively. We can build up the bases of $Clif(1, 5)$ using the following fact: If $(A_i, \nu_i)$ is a Clifford algebra for $\mathbb{R}^{t_i + s_i}$, then $(A_1 \otimes A_2, \nu)$ is a Clifford algebra for $\mathbb{R}^{t_1 + s_1} \oplus \mathbb{R}^{t_2 + s_2}$ where $\nu: (a_1, a_2) \mapsto a_1 \otimes 1_{A_2} + 1_{A_1} \otimes a_2$. This fact amounts to saying that, we can write down the gamma-matrices of the higher dimensional Clifford algebras by using the smaller ones under the following isomorphisms: 
\begin{align} 
Clif(t, s) \otimes Clif(1, 1) &\cong Clif(t+1, s+1), \nonumber \\ 
Clif(t, s) \otimes Clif(0, 2) &\cong Clif(s, t+2).
\label{e6}
\end{align}
In particular, we can write the gamma-matrices of $\mathbb{R}^{1+2}$ and $\mathbb{R}^{0+3}$ respectively as
\begin{align} 
\mathbb{R}^{1+2}: \quad \gamma^0 &= i \sigma_2,\ \gamma^1 = - \sigma_3, \ \gamma^2 = - \sigma_1, \nonumber \\ 
\mathbb{R}^{0+3}: \quad \gamma^5 &= - \tau_3, \  \gamma^6 = \tau_1, \ \gamma^7 = \tau_2.
\label{e7}
\end{align}
Here $\sigma_i, \tau_i$ and $\rho_i$ denote the $i$-th Pauli matrix
\begin{equation}
\sigma_1 := \tau_1 := \rho_1 := \left( \begin{array}{cc} 0 & 1 \\ 1 & 0 \end{array} \right), \qquad \sigma_2 := \tau_2 := \rho_2 := \left( \begin{array}{cc} 0 & -i \\ i & 0 \end{array} \right), \nonumber
\end{equation}
\begin{equation}
\sigma_3 := \tau_3 := \rho_3 := \left( \begin{array}{cc} 1 & 0 \\ 0 & -1 \end{array} \right),
\end{equation}
and they satisfy:
\begin{equation} \sigma_i \sigma_j = \delta_{i j} + i \epsilon_{i j k} \sigma_k,
\end{equation}
which is a particular example of the Clifford product (\ref{prod}). Clifford bundle of a manifold is locally the product of the manifold and $Clif(t, s)$, so on a coordinate chart the gamma-matrices of $\mathbb{M}^{1+2}$ and $\mathbb{S}^3$ are automatically the same as the ones of $\mathbb{R}^{1+2}$ and $\mathbb{R}^{0+3}$ respectively. Hence, we have the all necessary gamma-matrices in (\ref{e7}) and they satisfy the anti-commutation relations:
\[ \{ \gamma^{a}, \gamma^{b} \} = 2 \eta^{a b} \mathbb{I}_2, \qquad \qquad \{ \gamma^{\alpha}, \gamma^{\beta} \} = 2 \delta^{\alpha \beta} \mathbb{I}_2, \]
where $\mathbb{I}_n$ denotes $n \times n$ identity matrix, $\eta^{a b} = diag(- + +)$ and $\delta^{\alpha \beta}$ is the Kronecker delta symbol. By using the gamma-matrices that we constructed, we can write the gamma-matrices on $\mathbb{M}^{1+5}$ locally as follows
\begin{align} \Gamma^{a} &= \rho_1 \otimes \mathbb{I}_2 \otimes \gamma^{a}, \nonumber \\ 
							\Gamma^{\alpha} &= \rho_2 \otimes \gamma^{\alpha} \otimes \mathbb{I}_2,
\label{e8}
\end{align}
so that they satisfy $\{ \Gamma^A, \Gamma^B \} = 2 \eta^{A B} \mathbb{I}_8$. Here, we note that $\mathbb{M}^{1+5}$ has $8\times8$ gamma-matrices, whereas $\mathbb{M}^{1+2}$ and $\mathbb{S}^3$ have both $2\times2$ gamma-matrices. Due to this reason, in (\ref{e8}), we introduce $\rho_1$ and $\rho_2$ so that the dimensions match. \\

\noindent By using this decomposition, we can write the generators of the Lorentz algebra $\mathfrak{so}(1, 5)$ of $\mathbb{M}^{1+5}$ in terms of the gamma-matrices and generators of $\mathfrak{so}(1, 2)$ and $\mathfrak{so}(3)$:
\begin{align} \Sigma^{a b} &= \mathbb{I}_2 \otimes \mathbb{I}_2 \otimes \sigma^{a b}, \nonumber \\
							\Sigma^{a \beta} &= - \Sigma^{\beta a} = \frac{i}{2} \rho_3 \otimes \gamma^{\beta} \otimes \gamma^{a}, \nonumber \\
							\Sigma^{\alpha \beta} &= \mathbb{I}_2 \otimes \sigma^{\alpha \beta} \otimes \mathbb{I}_2,
\label{e9} 
\end{align}
where $\big\{ \Sigma^{A B} := \frac{1}{4} \left[ \Gamma^A, \Gamma^B \right] \big\}, \big\{ \sigma^{a b} := \frac{1}{4} \left[ \gamma^{a}, \gamma^{b} \right] \big\}$, and $\big\{ \sigma^{\alpha \beta} := \frac{1}{4} \left[ \gamma^{\alpha}, \gamma^{\beta} \right] \big\}$ are the Lorentz generators of $\mathbb{M}^{1+5}$, $\mathbb{M}^{1+2}$ and $\mathbb{S}^3$ respectively. \\

\noindent The spinor ansatz we make for the 6 dimensional Dirac equation is:
\begin{equation} \Psi \left( x^M \right) = \Psi(x^m, x^{\mu}) := \eta^i \otimes G(x^{\mu}) \xi^j \otimes \psi_{i j}(x^m),  \quad i, j = 1, 2
\label{e10.3}
\end{equation}
where $\{ \psi_{i j}(x^m) \}$ are spinor fields on $\mathbb{M}^{1+2}$ and $G(x^{\mu})$ is a generic $SU(2)$ element\footnote{For example, a generic spinor on $\mathbb{S}^3$ can be written with the help of Euler rotations of $SU(2)$:
\begin{equation} G(x^{\mu}) = G(\theta, \phi, \psi) := e^{\frac{i}{2} \tau_3 \phi} e^{\frac{i}{2} \tau_2 \theta} e^{\frac{i}{2} \tau_3 \psi}.
\end{equation}
where $\{ \theta, \phi, \psi \}$ are the hyper-spherical coordinates on $\mathbb{S}^3$. This can be done because $SU(2)$ is compact, so the exponential map $exp: \mathfrak{su}(2)\to SU(2)$ is surjective.}. Also, similar to (\ref{e8}), we introduce
\begin{equation} \eta^1 := \xi^1 := \begin{pmatrix} 1 \\ 0 \end{pmatrix}, \qquad \eta^2 := \xi^2 := \begin{pmatrix} 0 \\ 1 \end{pmatrix}, 
\end{equation}
in order to match the number of spinor components. Due to this reason, we have to introduce four different spinor fields $\{ \psi_{ij}(x^m) \ | \ i, j = 1, 2 \}$ on $\mathbb{M}^{1+2}$. \\

\noindent By using the spinor ansatz (\ref{e10.3}) and reduced forms of the quantities on $\mathbb{M}_6$ (\ref{e2}, \ref{e4}, \ref{e8} and \ref{e9}), we rewrite the Dirac equation (\ref{e10}) as 
\begin{align}
\star 1 \wedge \bigg\{\sigma_1\eta^i &\otimes G\xi^j \otimes \bigg[(\gamma^{c} *e_{c}) \wedge \bigg(d+\frac{1}{2}\omega_{a b}\sigma^{a b}\bigg)\bigg]\psi_{ij} \nonumber\\
+\sigma_1\eta^i &\otimes \bigg(-\frac{1}{4}\epsilon_{\alpha\beta}^{\ \ \ \gamma}\sigma^{\alpha\beta}\bigg) G\xi^j \otimes (A_{\gamma a}\gamma^{a}) \psi_{ij}*1 \nonumber\\
+\sigma_2\eta^i &\otimes \gamma^{\alpha}G\xi^j \otimes \bigg(\frac{1}{4}F_{\alpha a b}\sigma^{a b} \bigg) \psi_{ij} *1 \bigg\} \nonumber\\
+*1 \wedge \bigg\{ \sigma_1 \eta^i &\otimes \bigg[(A^{\gamma}_{\ a} \star e_{\gamma}) \wedge \bigg(d+\frac{1}{2}\omega_{\alpha\beta}\sigma^{\alpha\beta}\bigg)\bigg] G\xi^j \otimes \gamma^{a} \psi_{ij} \nonumber\\
-\sigma_2\eta^i &\otimes \bigg[(\gamma^{\gamma} \star e_{\gamma}) \wedge \bigg(d+\frac{1}{2}\omega_{\alpha\beta} \sigma^{\alpha\beta}\bigg)\bigg]G\xi^j \otimes \psi_{ij} \bigg\} \nonumber\\
= M \eta^i &\otimes G\xi^j\otimes \psi_{ij} *1 \wedge \star 1.
\label{rde}
\end{align}
We can get rid of the first factors in the triple tensor product by using the facts:
\begin{align}
\sigma_1 \eta^1 &= \eta^2, \qquad \qquad \ \sigma_1 \eta^2 = \eta^1, \nonumber \\
\sigma_2 \eta^1 &= i \eta^2, \qquad \qquad \sigma_2 \eta^2 = - i \eta^1.
\end{align}
This decomposes (\ref{rde}) into two equations that the spinors $\psi_{11}$ and $\psi_{12}$ of upper component get mixed with spinors $\psi_{21}$ and $\psi_{22}$ of lower component: 
\begin{align}
&\star 1 \wedge \bigg\{ G\xi^j \otimes \bigg[ (\gamma^{c}*e_{c}) \wedge \bigg(d+\frac{1}{2}\omega_{a b}\sigma^{a b} \bigg) \bigg] \psi_{2j} \nonumber\\
&\qquad+\gamma^{\alpha}G\xi^j \otimes \bigg(-\frac{i}{4}F_{\alpha a b}\sigma^{a b} \bigg) \psi_{2j} *1 \bigg\}  \nonumber\\
&+*1\wedge \bigg\{  \bigg[(A^{\gamma}_{\ a} \star e_{\gamma}) \wedge \bigg(d+\frac{1}{2}\omega_{\alpha\beta}\sigma^{\alpha\beta}\bigg)+\frac{1}{4}\epsilon_{\alpha\beta}^{\ \ \ \gamma}A_{\gamma a}\sigma^{\alpha\beta} \star 1 \bigg] G\xi^j \otimes \gamma^{a} \psi_{2j} \nonumber\\
&\qquad+ \bigg[i (\gamma^{\gamma} \star e_{\gamma}) \wedge \bigg(d+\frac{1}{2}\omega_{\alpha\beta} \sigma^{\alpha\beta}\bigg)\bigg]G\xi^j \otimes \psi_{2j} \bigg\} \nonumber\\
&=MG\xi^j \otimes \psi_{1j}*1 \wedge \star 1,
\label{rde1}
\end{align}
and
\begin{align}
&\star 1 \wedge \bigg\{ G\xi^j \otimes \bigg[ (\gamma^{c}*e_{c}) \wedge \bigg(d+\frac{1}{2}\omega_{a b}\sigma^{a b} \bigg) \bigg] \psi_{1j} \nonumber\\
&\qquad+\gamma^{\alpha}G\xi^j \otimes \bigg(\frac{i}{4}F_{\alpha a b}\sigma^{a b} \bigg) \psi_{1j} *1 \bigg\}  \nonumber\\
&+*1\wedge \bigg\{  \bigg[(A^{\gamma}_{\ a} \star e_{\gamma}) \wedge \bigg(d+\frac{1}{2}\omega_{\alpha\beta}\sigma^{\alpha\beta}\bigg)+\frac{1}{4}\epsilon_{\alpha\beta}^{\ \ \ \gamma}A_{\gamma a}\sigma^{\alpha\beta} \star1 \bigg] G\eta^j \otimes \gamma^{a} \psi_{1j} \nonumber\\
&\qquad+ \bigg[-i (\gamma^{\gamma} \star e_{\gamma}) \wedge \bigg(d+\frac{1}{2}\omega_{\alpha\beta} \sigma^{\alpha\beta}\bigg)\bigg]G\xi^j \otimes \psi_{1j} \bigg\} \nonumber\\
&=MG\xi^j \otimes \psi_{2j}*1 \wedge \star 1.
\label{rde2}
\end{align}
In these equations, we can see the Dirac operators on both $\mathbb{M}^{1+2}$ and $\mathbb{S}^3$:
\begin{align}
& \slashed{D}_{\mathbb{M}^{1+2}} := (\gamma^c * e_c) \wedge \left(d + \frac{1}{2} \omega_{a b} \sigma^{a b} \right), \qquad \mbox{acting on the spinors $\{ \psi_{i j} \}$,} \nonumber \\
& \slashed{D}_{\mathbb{S}^3} := (\gamma^{\gamma} \star e_{\gamma}) \wedge \left(d + \frac{1}{2} \omega_{\alpha \beta} \sigma^{\alpha \beta} \right), \qquad \mbox{\ acting on the spinors $\{ G \xi^j \}$.}
\end{align}
We also have a term similar to the Dirac operator on $\mathbb{S}^3$, but the gamma-matrix valued 1-form is of the form $A^{\gamma}_{\ a} e_{\gamma} \gamma^a$. The curvature of the gauge potential appears in the equations as a coupling with gamma-matrices of both $\mathbb{M}^{1+2}$ and $\mathbb{S}^3$:
\[ \frac{i}{4}F_{\alpha a b}\sigma^{a b} \gamma^{\alpha}. \]
Moreover, the gauge potential also couples to the equations minimally:
\[ \frac{1}{4}\epsilon_{\alpha\beta}^{\ \ \ \gamma}A_{\gamma a}\sigma^{\alpha\beta} \gamma^a. \]

\noindent For comparison, we also write the Dirac equation for a minimally coupled $SU(2)$ isodoublet on $\mathbb{M}^{1+2}$. An $SU(2)$ isodoublet is given by:
\begin{equation}
\tilde{\Psi}(x^m) := \tilde{G}(x^m) \xi^i \otimes \tilde{\psi}_{i}(x^m),  \qquad i = 1, 2,
\label{e15}
\end{equation}
where $\tilde{G}(x^m)$ is a generic local $SU(2)$ element as in the footnote 10. $\{ \tilde{\psi_{i}}(x^m) \}$ are 2-component spinor fields on $\mathbb{M}^{1+2}$ which are coupled minimally to the $\mathfrak{su}(2)$ valued potential 1-form $A := A_{\alpha} \gamma^{\alpha}$ through the Dirac operator: 
\begin{align}
\slashed{D}_A &:= *\gamma \wedge D_A \nonumber\\
&:= (\gamma^c*e_c) \wedge \bigg[ d (\mathbb{I}\otimes \mathbb{I}) +\frac{1}{2} \omega_{a b} (\mathbb{I} \otimes \sigma^{a b})+  A_{\alpha}(\gamma^{\alpha}\otimes \mathbb{I}) \bigg]\nonumber\\
&= (\gamma^c*e_c) \wedge \bigg[ (d+A)\otimes \mathbb{I} + \mathbb{I} \otimes \bigg(d+\frac{1}{2}\omega_{a b}\sigma^{a b}\bigg) \bigg].
\label{e16}
\end{align} 
The Dirac equation\footnote{It transforms covariantly under a local $SU(2)$ transformation $U=U(x^m)$ as:
\begin{align}
\tilde{\Psi} &\mapsto (U \otimes \mathbb{I})\Psi=U\tilde{G}\xi^i \otimes \tilde{\psi}_i, \nonumber\\
A \otimes \mathbb{I} &\mapsto (U \otimes \mathbb{I}) \big[(A+d)\otimes\mathbb{I}\big] (U^{-1} \otimes \mathbb{I})=U(A+d)U^{-1}\otimes \mathbb{I}.
\end{align}} satisfied by the isodoublet (\ref{e15}) is given by:
\begin{equation}
\slashed{D}_A \tilde{\Psi} = M*\tilde{\Psi}, \label{nade}
\end{equation}
and it explicitly reads:
\begin{equation}
(d+A) G\xi^i \otimes (\gamma^c*e_c) \psi_i + G\xi^i \otimes \slashed{D}_{\mathbb{M}^{1+2}} \psi_i = M G\xi^i \otimes \psi_i *1.
\end{equation}

\noindent To simplify the equations (\ref{rde1}) and (\ref{rde2}), we assume that the spinors on $\mathbb{S}^3$ are mass eigenstates of the Dirac operator on $\mathbb{S}^3$:
\begin{equation}
\slashed{D}_{\mathbb{S}^3} G \xi^j = m G \xi^j \star 1,
\end{equation}
for some mass term $m \in \mathbb{R}^{\geq 0}$. With this assumption, the equations (\ref{rde1}) and (\ref{rde2}) become
\begin{align}
&\star 1 \wedge \bigg\{ G\xi^j \otimes \bigg[ (\gamma^{c}*e_{c}) \wedge \bigg(d+\frac{1}{2}\omega_{a b}\sigma^{a b} \bigg) \bigg] \psi_{2j} \nonumber\\
&\qquad+\gamma^{\alpha}G\xi^j \otimes \bigg(-\frac{i}{4}F_{\alpha a b}\sigma^{a b} \bigg) \psi_{2j} *1 \bigg\}  \nonumber\\
&+*1\wedge \bigg\{  \bigg[(A^{\gamma}_{\ a} \star e_{\gamma}) \wedge \bigg(d+\frac{1}{2}\omega_{\alpha\beta}\sigma^{\alpha\beta}\bigg)+\frac{1}{4}\epsilon_{\alpha\beta}^{\ \ \ \gamma}A_{\gamma a}\sigma^{\alpha\beta} \star 1 \bigg] G\xi^j \otimes \gamma^{a} \psi_{2j} \nonumber\\
&\qquad+ i m G\xi^j \otimes \psi_{2j} \bigg\} \nonumber\\
&=MG\xi^j \otimes \psi_{1j}*1 \wedge \star 1,
\label{rde1b}
\end{align}
and
\begin{align}
&\star 1 \wedge \bigg\{ G\xi^j \otimes \bigg[ (\gamma^{c}*e_{c}) \wedge \bigg(d+\frac{1}{2}\omega_{a b}\sigma^{a b} \bigg) \bigg] \psi_{1j} \nonumber\\
&\qquad+\gamma^{\alpha}G\xi^j \otimes \bigg(\frac{i}{4}F_{\alpha a b}\sigma^{a b} \bigg) \psi_{1j} *1 \bigg\}  \nonumber\\
&+*1\wedge \bigg\{  \bigg[(A^{\gamma}_{\ a} \star e_{\gamma}) \wedge \bigg(d+\frac{1}{2}\omega_{\alpha\beta}\sigma^{\alpha\beta}\bigg)+\frac{1}{4}\epsilon_{\alpha\beta}^{\ \ \ \gamma}A_{\gamma a}\sigma^{\alpha\beta} \star 1 \bigg] G\eta^j \otimes \gamma^{a} \psi_{1j} \nonumber\\
&\qquad -i m G\xi^j \otimes \psi_{1j} \bigg\} \nonumber\\
&=MG\xi^j \otimes \psi_{2j}*1 \wedge \star 1.
\label{rde2b}
\end{align}

\noindent In order to observe free behavior and find the mass eigenstates of the Dirac operator on $\mathbb{M}^{1+5}$, we ignore the interaction by setting $A = F = 0$. Then, the equations (\ref{rde1}, \ref{rde2}) become
\begin{equation}
\xi^j \otimes \left( \slashed{D}_{\mathbb{M}^{1+2}} - i m * 1 \right) \psi_{2 j} = M \xi^j \otimes \psi_{1j} * 1,
\label{rde1me}
\end{equation}
and
\begin{equation}
\xi^j \otimes \left( \slashed{D}_{\mathbb{M}^{1+2}} + i m * 1 \right) \psi_{1 j} = M \xi^j \otimes \psi_{2j} * 1.
\label{rde2me}
\end{equation}
These equations mixes $\{ \psi_{1 j} \}$ with $\{ \psi_{2 j} \}$. In matrix form, we can write them as separate equations for $\{ \psi_{i 1} \}$ and $\{ \psi_{i 2} \}$:
\begin{equation}
\left( \begin{matrix} \slashed{D}_{\mathbb{M}^{1+2}} & 0 \\ 0 & \slashed{D}_{\mathbb{M}^{1+2}} \end{matrix} \right) \left( \begin{matrix} \psi_{11} \\ \psi_{21} \end{matrix} \right) = \left( \begin{matrix} M & i m \\ - i m & M \end{matrix} \right) \left( \begin{matrix} \psi_{11} \\ \psi_{21} \end{matrix} \right) * 1,
\end{equation}
and
\begin{equation}
\left( \begin{matrix} \slashed{D}_{\mathbb{M}^{1+2}} & 0 \\ 0 & \slashed{D}_{\mathbb{M}^{1+2}} \end{matrix} \right) \left( \begin{matrix} \psi_{12} \\ \psi_{22} \end{matrix} \right) = \left( \begin{matrix} M & i m \\ - i m & M \end{matrix} \right) \left( \begin{matrix} \psi_{12} \\ \psi_{22} \end{matrix} \right) * 1.
\end{equation}
The mass matrix $\left( \begin{matrix} M & i m \\ - i m & M \end{matrix} \right)$ has eigenvalues $M \pm m$ corresponding to mass eigenspinors 
\[ \psi_{\pm i} := \frac{1}{2} \left( \psi_{1 i} \pm \psi_{2 i} \right). \]
This shows that in order to have non-negative masses we should assume $M \geq m$. When $m = 0$, the mass matrix is already diagonal and mass eigenspinors on compact $\mathbb{S}^3$ corresponds to zero modes of the Dirac operator. 


\section{Concluding Remarks}

\noindent In this work, a dimensional reduction of the Dirac equation of $\mathbb{M}^{1+5} = \mathbb{M}^{1+2} \times \mathbb{S}^3$ on $\mathbb{S}^3$ is examined. Mathematical structures that one encounters for the Dirac equation and Kaluza-Klein reduction are discussed in detail. Throughout the paper, we aimed to follow a pedagogical path, so we tried to point out important definitions and facts about these structures and our calculations. \\

\noindent The Dirac equation (\ref{e10}) explains the motion of free spinor fields on $\mathbb{M}^{1+5}$. The compact part $\mathbb{S}^3$ has a $SU(2)$ group structure, so $\mathbb{M}^{1+5}$ can be seen as a principal $SU(2)$ bundle over $\mathbb{R}^{1 + 2}$. This bundle structure allows one to perform a Kaluza-Klein reduction on the equation and reduce it to several equations on smaller parts of the spacetime. After the reduction, one gets the Dirac terms on both $\mathbb{R}^{1 + 2}$ and $\mathbb{S}^3$ with some additional interaction terms. These interactions are inherited from the group structure of $\mathbb{S}^3$, so they contain minimal and non-minimal couplings of the non-abelian $SU(2)$ Yang-Mills potentials. We also compared these couplings with the usual minimal $SU(2)$ coupling on $\mathbb{M}^{1+2}$. \\

\noindent Topological insulators became a really important topic after the recent developments in both theoretical and experimental fields. Free topological insulators are classified completely by using several different methods \cite{kitaev}, \cite{ryu-etal}. Some of these techniques are even fruitful for the interacting ones \cite{ryu-takayanagi}, \cite{ryu-moore-ludwig}. An example of non-abelian interactions of topological insulators with artificial gauge potentials are studied in \cite{hauke-etal}. We believe that our calculations about minimal and non-minimal $SU(2)$ couplings of the reduced theory may help us to understand these kind of interactions between the topological insulators and gauge fields. Moreover, we hope this work will be useful for the classification of interacting topological insulators. \\


\end{document}